\documentclass[sigconf]{acmart}

\copyrightyear{2026}
\acmYear{2026}
\setcopyright{cc}
\setcctype{by-nc-nd}
\acmConference[UIST '26]{The 39th Annual ACM Symposium on User Interface Software and Technology}{November 02--05, 2026}{Detroit, MI, USA}
\acmBooktitle{The 39th Annual ACM Symposium on User Interface Software and Technology (UIST '26), November 02--05, 2026, Detroit, MI, USA}
\acmDOI{10.1145/3830398.3830486}
\acmISBN{979-8-4007-2856-3/2026/11}

\usepackage{enumitem}
\usepackage{graphicx}
\usepackage{multirow}

\newcommand{\final}[1]{\textcolor{black}{#1}}

\newcommand{\rev}[1]{\textcolor{black}{#1}}

\AtBeginDocument{%
  }

\newcommand{\eg}{{\it e.g.,\ }}

\newcommand{\ie}{{\it i.e.,\ }}
\newcommand{\tool}{{Sonic Stage}~}
\newcommand{\toole}{{Sonic Stage}}

\begin{document}

\title[\toole]{\toole: \rev{Auto-Generating} Interactive Spatial Soundscapes to Facilitate Dialogue Video Comprehension for Blind Viewers}

\author{Shuchang Xu}
\affiliation{
\institution{Hong Kong University of Science and Technology}
\city{Hong Kong}
\country{China}
}
\orcid{0000-0002-7642-9044}
\email{sxuby@connect.ust.hk}

\author{Xiaofu Jin}
\affiliation{
\institution{Univeristy of Stuttgart}
\city{Stuttgart}
\country{Germany}
}
\orcid{0000-0002-7239-3769}
\email{xiaofu.jin@vis.uni-stuttgart.de}

\author{Gaurav Jain}
\affiliation{
\institution{Columbia University}
\city{New York}
\country{USA}
}
\orcid{0000-0002-9084-9395}
\email{gaurav@cs.columbia.edu}

\author{Wenshuo Zhang}
\affiliation{
\institution{Hong Kong University of Science and Technology}
\city{Hong Kong}
\country{China}
}
\orcid{0009-0007-9226-0713}
\email{wzhangeb@connect.ust.hk}

\author{Huamin Qu}
\affiliation{
\institution{Hong Kong University of Science and Technology}
\city{Hong Kong}
\country{China}
}
\orcid{0000-0002-3344-9694}
\email{huamin@cse.ust.hk}

\author{Brian A. Smith}
\affiliation{
\institution{Columbia University}
\city{New York}
\country{USA}
}
\orcid{0000-0003-2540-0839}
\email{brian@cs.columbia.edu}

\author{Yukang Yan}
\authornote{This is the corresponding author.}
\affiliation{
\department{Department of Computer Science}
\institution{University of Rochester}
\city{New York}
\country{United States}
}
\orcid{0000-0001-7515-3755}
\email{yukang.yan@rochester.edu}

\renewcommand{\shortauthors}{Xu et al.}

\begin{abstract}
Audio description (AD) makes film and television accessible to blind and low-vision (BLV) audiences by narrating characters' actions. 
However, in scenes with lots of dialogue, AD often omits important actions because it is constrained not to overlap with speech. 
It is not yet known how to convey characters' actions during dialogue. 
We present \toole, a system that transforms dialogue videos into \textit{interactive spatial soundscapes}, enabling BLV audiences to intuitively understand characters' actions and movements through immersive auditory cues. 
\tool conveys essential visual information during dialogue through three auditory techniques: (1) \textit{spatialized dialogue} to represent spatial layout, (2) \textit{diegetic sound} to convey character actions, and (3) \textit{interactive descriptions} to provide context-specific visual details. Evaluation with 12 BLV viewers showed that \tool significantly improved video comprehension, spatial presence, and narrative engagement. 
We highlight opportunities for enhancing video accessibility across diverse genres through immersive, interactive audio representations.
\end{abstract}



\begin{teaserfigure}
  \centering
  \includegraphics[width=1.0\textwidth]{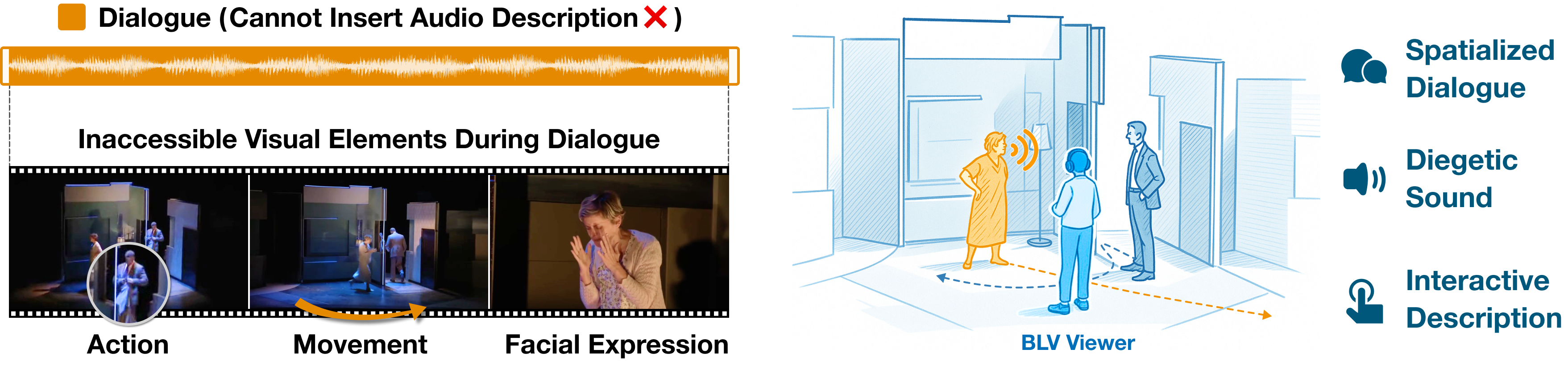}
  \caption{
  (Left) In videos with lots of dialogue, there is little opportunity to insert audio descriptions. 
  As a result, blind viewers often miss crucial visual information, such as characters' actions and movements. 
  (Right) \tool conveys essential visual information during dialogue through three auditory techniques: \textit{spatialized dialogue}, \textit{diegetic sound}, and \textit{interactive descriptions}. 
  These techniques enable BLV viewers to perceive on-screen actions within an immersive auditory experience.
  }
  \label{fig:teaser}
\end{teaserfigure}

\maketitle

\section{Introduction}

Movies and television (TV) shows are powerful storytelling media that combine rich visual and auditory elements to create immersive experiences \cite{visch2010emotional,bracken2010immersed}. 
Yet, their visual content remains largely inaccessible to blind and low vision (BLV) viewers. 
The primary approach for making such videos accessible is to provide audio description (AD) for the visual content \cite{amy2020rescribe}. 
AD narrates key visual elements such as characters' actions and expressions during natural gaps in speech \cite{amy2020rescribe,wang2021Tiresias}. 
For AD to work effectively, long gaps in speech are needed to afford enough time for inserting descriptions. 
In scenes with dense dialogue, however, there is little opportunity to insert descriptions \cite{amy2020rescribe,asset2024customAD}, making AD much less effective. 
For example, Figure~\ref{fig:teaser} (left) shows two characters engaging in rich actions, movements, and facial expressions, yet their continuous conversation leaves little space to describe these details. 
Consequently, BLV viewers often miss crucial on-screen actions that occur during dialogue. 
This highlights the need for approaches that can complement AD and effectively convey visual information during speech.

\rev{Prior work has explored conveying visual information through non-verbal audio \cite{jain2023frontrow,jiang2023_beyond_AD,lopez2021enhancing}. 
For instance, spatial audio has been used in tennis broadcasts to indicate player positions and movements \cite{jain2023frontrow}. 
However, unlike sports, films and TV shows typically lack a consistent spatial representation, creating unique challenges for designing spatial audio cues. 
Other approaches have enabled BLV viewers to access video content through touch-based exploration \cite{ning2024spica} or visual question answering \cite{stangl2023potential}. 
While these methods enhance user agency, 
they often disrupt the video's narrative flow and reduce viewer immersion \cite{ning2024spica,cheema2025describe}. 
Consequently, making dialogue videos accessible to BLV audiences while preserving an immersive viewing experience remains an open challenge.}

To address this challenge, we present \toole, a system that automatically transforms dialogue videos into interactive spatial soundscapes, 
enabling BLV viewers to intuitively understand characters' actions and movements through immersive auditory cues.
\tool conveys essential visual information during dialogue through three core techniques. 
First, \textbf{spatialized dialogue} represents the spatial layout of characters within a scene, allowing viewers to perceive characters' positions and movements via spatial audio. 
Second, \textbf{diegetic sound} conveys characters' actions through non-verbal sound effects that reflect the on-screen actions. 
Third, \textbf{interactive descriptions} provide context-specific visual details related to the dialogue, which viewers can access on demand through touchscreen input. 
\final{A key novelty of \tool is its ability to preserve spatial-audio coherence across camera cuts. 
In film and television, camera cuts can abruptly change the on-screen positions of characters, 
causing screen-anchored audio cues to shift suddenly and disorient BLV viewers. 
To address this issue, \tool reconstructs a stable \textbf{3D scene representation} across shots and renders all auditory cues within a shared spatial reference frame, thereby maintaining a coherent auditory experience across camera cuts.}

To evaluate \toole, we conducted a within-subject study with 12 BLV participants, who compared \tool with a baseline modeled after SPICA \cite{ning2024spica}, 
\rev{the state-of-the-art system that provides both spatial and semantic access to video content.} 
Each participant viewed two matched sets of dialogue videos using the two systems, respectively. 
Results showed that \tool significantly improved video comprehension ($p<.01$) in both spatial and semantic aspects. 
Participants' subjective ratings indicated higher spatial presence ($p < .01$), narrative engagement ($p < .01$), overall enjoyment ($p < .01$), and willingness to use the system in the future ($p < .01$). 
Participants noted that \tool made videos feel more \textit{expressive} and \textit{engaging}, 
and they suggested extending its application to additional genres such as documentaries, musical theater, and dance performances. 
Based on these findings, we discuss opportunities for more immersive, interactive audio representations to advance video accessibility for BLV audiences.

Our contributions are threefold:
\begin{itemize}[leftmargin=*, labelindent=0pt, itemindent=0pt]
    \item \rev{We introduce three auditory techniques --- \textit{spatialized dialogue}, \textit{diegetic sound}, and \textit{interactive descriptions} --- 
    that complement AD and convey essential visual information during speech segments.}
    \item \rev{We present an automated pipeline that reconstructs a \textit{3D spatial soundscape} from cross-cut videos, thereby providing spatially coherent auditory cues for BLV audiences.}
    \item We contribute an evaluation study that demonstrates how BLV viewers consume dialogue videos using \tool and highlight opportunities for immersive, interactive audio representations to enhance video accessibility across diverse genres.
\end{itemize}

\section{Related Work}
Our work builds on two core research areas: 
(1) video accessibility for BLV viewers, and 
(2) spatial audio generation from videos.

\subsection{Video Accessibility for BLV Viewers}
\rev{Videos are often inaccessible to BLV viewers when the visual content cannot be understood through audio alone \cite{van2024making,liu2021makes}. 
To make videos accessible to BLV audiences, prior work has explored various approaches, including audio description, non-verbal audio, haptic feedback, and interactive systems.}

\subsubsection{\textbf{Audio Description}} 
Audio description (AD) is the primary approach for making videos accessible to BLV audiences \cite{amy2020rescribe,wang2021Tiresias}. 
AD narrates key visual elements (\eg characters' actions, facial expressions, and scene settings) during natural gaps in dialogue. 
\rev{
Prior work \cite{amy2020rescribe} has established key guidelines for creating effective AD: 
(1) Describe essential visual elements not conveyed through audio, 
(2) Do not overlap AD with the video's dialogue or important sounds, and 
(3) Use an objective tone and avoid subjective interpretation. 
}
To produce effective AD, a core challenge lies in conveying key visual information within limited non-speech gaps \cite{amy2020rescribe,wang2021Tiresias}. 
To address this challenge, researchers have proposed manual \cite{killough2023exploring,liu2025cosight}, semi-automated \cite{amy2020rescribe,liu2022crossa11y,natalie2023supporting,cheema2025describepro}, and fully automated approaches \cite{videoa11y,wang2021Tiresias,xu2025danmua11y} for optimizing the content and placement of AD. 
For example, CrossA11y \cite{liu2022crossa11y} assists AD creators in identifying inaccessible visual content by detecting audio-visual discrepancies, 
while Rescribe \cite{amy2020rescribe} uses dynamic programming to help describers select which visual details to include within limited gaps. 
\rev{
When key visual information cannot fit within existing gaps, the Web Accessibility Initiative suggests using extended AD \cite{caldwell2008web}, which pauses the video to insert additional narration. 
While effective in educational videos \cite{peng2021slidecho}, extended AD has been shown to disrupt narrative coherence in film and TV shows \cite{jiangchi24context}. 
This limitation has motivated alternative approaches that convey visual information during dialogue, such as through non-verbal audio.
}

\subsubsection{\textbf{Non-Verbal Audio}} 

Non-verbal audio has been explored for enhancing video accessibility in different contexts, including sports broadcasts \cite{jain2023frontrow}, 360° videos \cite{jiang2023_beyond_AD}, and films \cite{lopez2021enhancing,lopez2022seeing}. 
For example, \textit{Front Row} \cite{jain2023frontrow} used spatial audio to represent player positions and movements in tennis matches, while Jiang et al. \cite{jiang2023_beyond_AD} explored diegetic sounds to subtly guide viewers’ attention in 360° videos. 
\final{
While these studies demonstrate the potential of spatial audio for video accessibility, they typically rely on a predefined spatial layout. However, film and television lack a consistent spatial representation across camera shots. To address this limitation, \tool reconstructs a shared 3D scene representation and renders the auditory cues within this spatial frame, thereby providing BLV viewers with coherent spatial information across camera cuts.
}

\subsubsection{\textbf{Haptic Feedback}} 
\rev{
In addition to auditory cues, prior work has explored conveying visual information in videos through haptic feedback \cite{viswanathan2010haptics,mcdaniel2013evaluation,jiangchi24context}. 
For example, Viswanathan et al. \cite{viswanathan2010haptics} developed haptic belts that used vibrations to signal characters' positions in films, 
while Jiang et al. \cite{jiangchi24context} proposed using tactile graphics to make tutorial videos accessible. 
Beyond videos, haptic feedback has been explored to improve the accessibility of images \cite{mackowski2023multimodal,butler2021technology,holloway2018accessible,bliss2007optical}, virtual reality \cite{zhao2018enabling,tzovaras2004design}, and real-world environments \cite{xu2020virtual,liu2021tactile}. 
However, most haptic approaches rely on specialized hardware such as vibration motors \cite{liu2021tactile}, 
tactile graphics \cite{mackowski2023multimodal}, 
and tangible 3D models \cite{bleau2023cognitive}, 
which limits their scalability for widespread adoption \cite{jiang2025can}. 
Consequently, our work focuses on auditory cues that can be delivered through standard headphones without additional equipment.
}

\subsubsection{\textbf{Interactive Systems}}\label{sec:rw_systems}
Prior work has developed interactive systems that enable BLV viewers to explore video content in greater detail \cite{asset2024customAD,xu2025branch,van2024making,ning2024spica,stangl2023potential,peng2021slidecho,cheema2025describe,xu2024memory}. 
These systems primarily increase user agency by allowing viewers to request additional information on demand. 
For instance, ShortScribe \cite{van2024making} enables users to access video descriptions at multiple levels of granularity, 
while Describe Now \cite{cheema2025describe} allows viewers to pause playback and request frame-level descriptions. 
InfoBot \cite{stangl2023potential} further enables viewers to ask visual questions about paused frames. 
While these systems enhance user agency, they often disrupt the video flow \cite{ning2024spica,cheema2025describe}, and how to maintain narrative coherence during video exploration remains an open challenge. 
To address this challenge, \tool introduces three auditory techniques that convey visual information during dialogue, thereby minimizing disruption to the viewing experience.

Another key distinction lies in the type of information delivered to users. 
While existing systems primarily convey semantic information, \tool provides both semantic and spatial information. 
The system most closely related to ours is SPICA \cite{ning2024spica}, which supports spatial and semantic access through two interaction modes: (1) \textit{spatial exploration}, in which users drag a finger across a paused frame to hear object descriptions, and (2) \textit{temporal exploration}, in which users navigate between key frames to access associated descriptions. Our comparative study between \tool and SPICA showed that directly integrating spatial information into the soundtrack improved spatial comprehension and immersion for BLV audiences.

\subsection{Spatial Audio Generation from Videos}
Generating spatial audio from mono-audio videos is a challenging task that involves three main steps \cite{lin2023soundify,morgado2018self}: \textit{separating} sound sources, \textit{localizing} them within the frame, and \textit{aligning} their positions across frames. Prior research has explored manual \cite{ccamci2017inviso}, semi-automated \cite{lin2023soundify, mimosa_spatial_audio}, and fully automated approaches \cite{morgado2018self, xu2021visually, zhou2020sep} to generate spatial audio from videos. 
For instance, Soundify \cite{lin2023soundify} adjusts the panning and volume of sound to create spatial audio, 
while Sep-Stereo \cite{zhou2020sep} employs deep neural networks to generate spatial audio based on object locations in the video frame. 
However, most existing approaches generate spatial audio in the \textit{screen space}, which can cause abrupt changes in sound location when the camera view shifts \cite{chion2019audio}. 
To address this limitation, \tool generates spatial audio within the \textit{3D scene space}, making it suitable for multi-view videos recorded within the same environment.

Reconstructing 3D scenes from film or TV clips is especially challenging due to sparse camera views and diverse shot types \cite{liu2022depth}. Traditional techniques, including structure-from-motion \cite{schonberger2016structure, hartley2003multiple, cui2017hsfm} and 3D Gaussian splatting \cite{kerbl20233d, qin2024langsplat}, face significant limitations under sparse-view conditions. To address this challenge, we leverage the Visual Geometry Grounded Transformer (VGGT) \cite{wang2025vggt}, 
which employs a feed-forward neural network to directly reconstruct 3D point clouds from sparse views. 
We further use a frame sampling strategy that selects key frames with meaningful spatial context, enabling coherent spatial audio generation across shot changes.

\section{\toole}

\tool is a system that transforms dialogue videos into \textit{interactive spatial soundscapes}, enabling BLV viewers to perceive characters' actions and movements through immersive audio cues. 
It conveys essential visual information during dialogue through three core techniques: 
(1) \textit{spatialized dialogue} to represent spatial layout, 
(2) \textit{diegetic sound} to convey character actions, and 
(3) \textit{interactive descriptions} to provide context-specific visual details. 
These techniques address key information needs identified in prior studies \cite{lopez2021enhancing, jiang2023_beyond_AD,jiangchi24context} and AD guidelines \cite{netflix2024ADStyleGuide,videoa11y,Ofcom2024AccessServicesGuidelines}. 
All auditory cues are rendered within a \textit{consistent 3D soundscape}, ensuring spatial continuity across camera view changes \cite{chion2019audio,sonic_stage_chi26ea}. 
In the following, we first present a walkthrough of the system (Section~\ref{sec:walkthrough}), followed by its audio generation pipeline (Sections~\ref{sec:pipeline_overview}--\ref{sec:module3}).

\subsection{System Walkthrough}\label{sec:walkthrough}
\begin{figure*}[h]
    \centering
    \includegraphics[width=1.0\linewidth]{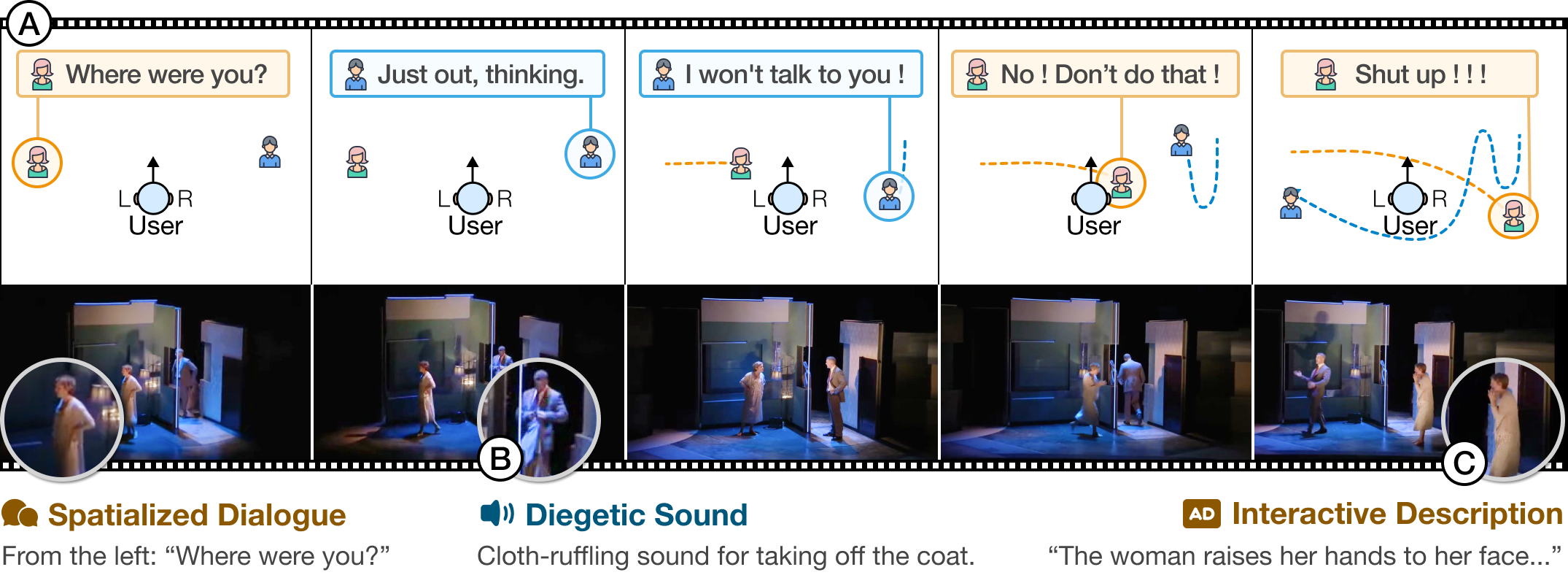}
    \caption{A walkthrough of \toole. (A) Throughout the video, users hear \textbf{spatialized dialogue} coming from the speakers' locations in the scene. 
    (B) When a character performs an action, users hear a \textbf{diegetic sound} from the location of that action. 
    (C) Users can access context-specific visual details through \textbf{interactive descriptions}. 
    For example, when the woman shouts ``\textit{Shut up!}'', a single tap on the phone plays the description: ``\textit{The woman raises her hand to her face and shouts.}''.}
    \label{fig:4_1_walkthrough}
    \Description{This figure provides a walkthrough of Sonic Stage, showing how users experience different auditory techniques. (A) Spatialized dialogue helps users perceive character positions, such as recognizing the woman speaking “Where were you?” from the left side. (B) Diegetic sound conveys character actions, for instance, a cloth-ruffling sound indicates the man removing his coat at a specific location. (C) Interactive descriptions deliver context-specific visual details, such as “The woman raises her hand to her face and shouts,” when the character exclaims “Shut up!”. Together, these techniques allow BLV users to intuitively follow characters' actions, movements, and expressive gestures through an immersive soundscape.}
\end{figure*}

To demonstrate the system, we follow Jason, a blind viewer, as he watches a video of a couple arguing at home\footnote{The demo is provided in the video figure.}. \tool employs three auditory techniques to help Jason intuitively understand characters' actions and movements within the scene.

\textbf{Scene Preview.} 
Before the video begins, Jason hears an audio introduction generated by \tool that describes the setting and characters: ``\textit{You are in a dimly lit living room. To your left, a woman waits at home. To your right, a man has just returned.}'' This preview helps Jason form an initial overview of the scene.

\textbf{Spatialized Dialogue.} 
Throughout the video, Jason can hear each line of dialogue emanating from the speakers' position within the scene. 
Jason is virtually placed at the center of the room, between the two speakers in this case, as shown in Figure~\ref{fig:4_1_walkthrough}A. 
This spatialization enables him to perceive speakers' positions and movements. 
For example, Jason initially hears the woman's voice emanating from the left, allowing him to recognize her \textit{position} on that side. When the argument intensifies, the woman's voice rapidly shifts from left to right, and this quick panning enables Jason to perceive her \textit{movement} and infer her heightened anger.

\textbf{Diegetic Sound.} 
As the video progresses, Jason hears a cloth-ruffling sound at the man's location in the spatial soundscape (see Figure~\ref{fig:4_1_walkthrough}B). 
Curious about its meaning, he taps on the screen to hear ``\textit{The man takes off his coat and walks into the room}''. 
Sonic Stage conveys such character actions through diegetic sound cues in the soundscape. 
These cues directly represent the action itself --- \eg a cloth-ruffling sound for removing a coat or a page-flipping sound for reading a book --- allowing Jason to perceive character actions intuitively without interrupting the dialogue.

\textbf{Interactive Description.} 
Later in the scene, when the woman exclaims, ``\textit{Shut up!}'', Jason becomes curious about her appearance. With a single tap on the screen, he hears: ``\textit{The woman raises her hands to her face and shouts.}'' 
This description allows Jason to perceive the woman’s desperation more vividly. 
\tool provides such descriptions that emphasize the current speaker's appearance and referenced objects,  giving Jason access to the most relevant details with minimal disruption. 
Once the description ends, the video auto-resumes, allowing Jason to re-enter the narrative seamlessly.

\subsection{Pipeline Overview}\label{sec:pipeline_overview}
\toole's pipeline generates the interactive spatial soundscapes through three core modules: 
(1) \textit{Spatialized Dialogue Generation}, which uses 3D reconstruction and optimization techniques to spatialize character dialogue within the scene space. 
(2) \textit{Diegetic Sound Generation}, which employs text-to-sound models to generate diegetic sound for key character actions. 
(3) \textit{Interactive Description Generation}, which uses multimodal language models to describe visual details relevant to the dialogue. 
This pipeline is designed for dialogue scenes where multiple characters interact within the same space, a common setting in drama, comedy, and film \cite{leake2017dialogue,girmaji2025dialogue}. Extensions to other settings are discussed in Section~\ref{sec:future_work}.

\subsection{Spatialized Dialogue Generation}\label{sec:module1}
\tool spatializes the dialogue within a consistent 3D soundscape through three key steps (see Figure~\ref{fig:module1}): 
(1) \textit{frame sampling}, 
(2) \textit{scene reconstruction}, and 
(3) \textit{soundscape optimization}.

\subsubsection{\textbf{Frame Sampling}}

To enable 3D reconstruction, the system samples frames based on two factors: \textit{spatial informativeness} and \textit{motion richness}. 
For {spatial informativeness}, the system selects only full and medium shots, 
because they capture both characters and the environment in balance \cite{shot_size}. 
Shot boundaries are detected using SceneDetect~\cite{scenedetect} and shot sizes are classified with Qwen-2-VL~\cite{bai2025qwen2}, a state-of-the-art model for shot-type recognition ~\cite{liu2025shotbench}. 
For {motion richness}, the system identifies moving characters by tracking their bounding box centroids, identified by YOLOv11 \cite{khanam2024yolov11}. 
A character is considered moving if its displacement per second exceeds 25\% of its bounding-box width (horizontal) or height (vertical). 
For shots containing moving characters, frames are sampled once per second; otherwise, only the middle frame is sampled.

\subsubsection{\textbf{Scene Reconstruction}}\label{sec:scene_reconstruction}
All sampled frames are processed by VGGT~\cite{wang2025vggt} to reconstruct 3D point clouds and recover character trajectories  (see Figure~\ref{fig:module1}(B)). 
Characters are detected in each frame using YOLOv11~\cite{khanam2024yolov11} and tracked across frames with BoT-SORT~\cite{aharon2022bot}. 
For each frame, the 2D segmentation mask of a character is projected onto the VGGT-generated 3D point cloud. The character’s 3D position is estimated as the mean coordinate of ten randomly sampled points within the masked region. 
After all frames are processed, character positions are interpolated at one-second intervals to form trajectories. 
For shots without motion, characters are assigned the position of the middle frame. Any remaining gaps are filled using linear interpolation to produce smooth trajectories.

\subsubsection{\textbf{Soundscape Optimization}}
\begin{figure}[!t]
    \centering
    \includegraphics[width=1.0\columnwidth]{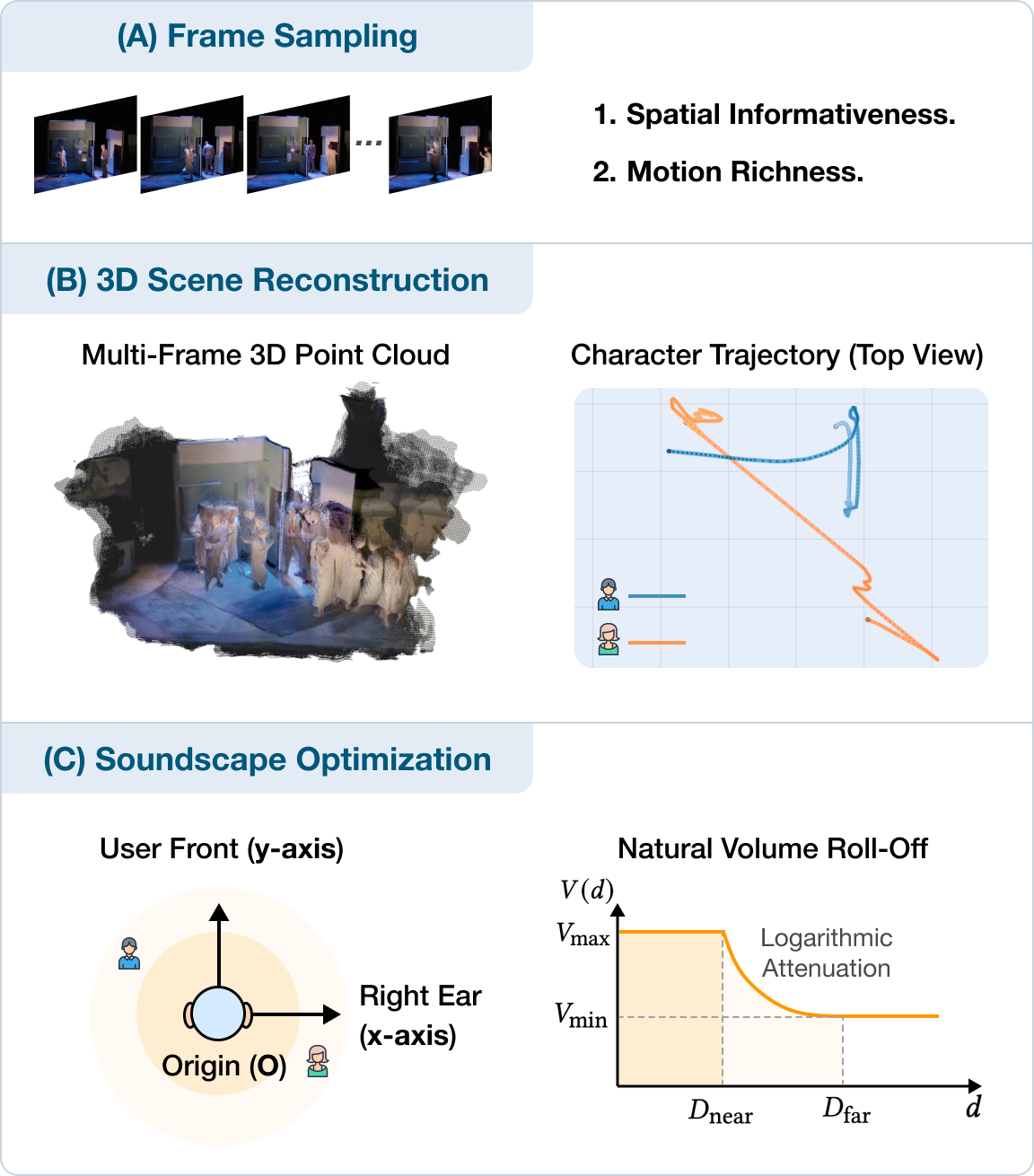}
    \caption{
    \tool constructs a coherent spatial soundscape in three stages: (A) selecting frames with high \textit{spatial} and \textit{motion} information; (B) reconstructing 3D point clouds from sampled frames to recover character trajectories; and (C) mapping these trajectories onto an optimized soundscape to enhance spatial distinction and model natural roll-off.}
    \label{fig:module1}
    \Description{This figure illustrates Module 1 of Sonic Stage, which generates spatialized dialogue through three steps. (A) Frame sampling selects frames based on spatial informativeness (favoring full and medium shots) and motion richness (adapting sample rates to movements). (B) Scene reconstruction builds 3D point clouds from sampled frames and recovers character trajectories from a top-down view. (C) Soundscape optimization maps these trajectories to an optimized soundscape, enhancing left–right distinction and applying natural volume roll-off to simulate realistic sound attenuation.}
\end{figure}

Given the characters' trajectories, we map them onto a 3D soundscape that preserves \textit{distinguishable} spatial cues while maintaining \textit{smooth} audio transitions. 
As shown in Figure~\ref{fig:module1} (C), we need to determine three elements: the listener’s position (the origin \({O}\)), the listener’s left–right ear orientation (the \({x}\){-axis}), and a volume roll-off function \(V(d)\). 
We derive these elements through the following optimization process.

\textbf{Distinguishing positions and movements.} 
Humans are most sensitive to left–right auditory differences \cite{cho2024auptimize}. 
We therefore align the \({x}\)-axis with the direction of greatest spatial variance among speakers and place the listener at their geometric center. 
Formally, given character positions \(\mathbf{p}_i^t \in \mathbb{R}^2\), where $i$ indexes characters and $t$ denotes time (sampled once per second), we have:

\[
\mathbf{x} = \arg\max_{\|\mathbf{v}\|=1} \sum_{t=1}^{T} \sum_{i=1}^{N} \left( \mathbf{v}^\top \mathbf{p}_i^t \right)^2, \quad \mathbf{O} = \frac{1}{TN} \sum_{t=1}^{T} \sum_{i=1}^{N} \mathbf{p}_i^t 
\]

\textbf{Conveying distance changes.} 
To convey speaker distance changes, we modulate speech volume as a function of the speaker--listener distance $d$. 
We adopt a logarithmic attenuation curve that simulates real-world volume roll-off \cite{mellow2012acoustics} (see Figure~\ref{fig:module1} (C)):

\[
V(d) = V_{\max} - (V_{\max} - V_{\min})\,
\times
\max\!\Bigg(0,\; 
  \min\!\Big(1,\;
    \frac{\log (d/D_{\text{near}})}{\log (D_{\text{far}}/D_{\text{near}})}
  \Big)
\Bigg)
\]

Based on iterative testing with two professional BLV sound designers (SD1 and SD2; Table~\ref{tab:demographics}), we set $V_{\max} = 1.0$ and $V_{\min} = 0.5$ to preserve audibility at long ranges. 
We chose $D_{\text{near}}$ as the median and $D_{\text{far}}$ as the 90th percentile of all speaker–listener distances in the scene, thereby limiting the influence of outliers.

\textbf{Ensuring smooth spatial transitions.} 
To stabilize spatial audio cues, we smooth each character’s trajectory using a two-second moving average filter. 
To reduce abrupt directional changes during speaker transitions, spatial blending \cite{unitySpatialBlend} is applied with a factor of 0.7, such that that 70\% of the audio is rendered spatially and 30\% is rendered in mono. 
This configuration, validated by BLV sound designers (SD1 and SD2; Table~\ref{tab:demographics}), maintains clear spatial separation while reducing the distracting shifts reported in prior work \cite{jiang2023_beyond_AD,chang2022omniscribe}.

\subsubsection{\textbf{Dialogue Spatialization}}
The system spatializes characters' dialogue within the soundscape as follows. 
First, it separates dialogue from background music using the Gaudio Studio’s AI stem splitter \cite{stem_splitter}. 
The dialogue is then transcribed via the Tencent Transcription API \cite{tencent_transcription}, producing speaker-labeled transcripts with millisecond-level timestamps (\eg ``\textit{00:00:01,010 → 00:00:03,100 [Speaker 1] Look at this!}''). 
Speaker labels are linked to on-screen characters using TalkNet \cite{tao2021someone}. 
Finally, each speech segment is spatialized according to the speaker's position in the soundscape.

\subsection{Diegetic Sound Generation}\label{sec:module2}
\final{
Diegetic sounds can help BLV viewers interpret key actions \cite{lopez2021enhancing}. 
Yet, sounds associated with visible actions may be absent from the original soundtrack. 
Sonic Stage generates these missing sounds through a three-stage pipeline that 
(1) detects actions with missing sounds, 
(2) generates a diegetic sound for each action, and 
(3) aligns the generated sound with the corresponding action.
}

\textbf{\final{Missing Sound Detection.}}
\final{
To identify actions with missing sounds, the system uses Gemini-2.5 \cite{gemini_video}, which supports sound and action recognition from video inputs. 
For each video segment, the model receives the corresponding video clip and is prompted to identify character actions that (a) would be expected to produce distinct, non-vocal sounds (\eg \textit{opening a door}) and (b) lack the corresponding sounds in the original audio.} 
For each action with missing sounds, the model outputs (i) the action's start and end timestamps and (ii) a description of the expected sound (\eg \textit{a door opening forcefully in frustration}).

\textbf{Sound Generation.}
The system uses ElevenLabs's text-to-sound model to generate five candidate sounds per action. The sound duration is set to the detected action span. 
Each candidate is evaluated by Gemini-2.5 with audio understanding capabilities \cite{gemini_audio}. 
The model is instructed to 
``\textit{rate the similarity of the sound to the sound description on a scale of 1 to 10.}'' 
The candidate with the highest score is selected as the final sound.

\textbf{Sound--Action Alignment.} 
The selected sound is played at the detected start time of the corresponding action and spatialized according to the action's location in the scene. 
At the onset of the sound, the playback device (\ie, a phone) also produces a half-second vibration. This haptic cue informs users that interactive descriptions associated with the action are available.

\subsection{Interactive Description Generation}\label{sec:module3}
\final{
Prior studies have shown that BLV viewers value on-demand access to visual details \cite{ning2024spica,cheema2025describe}. 
Yet, pausing playback to provide additional information can disrupt narrative flow. 
To minimize such disruption while preserving user agency, 
\tool generates \textit{concise}, \textit{dialogue-relevant} descriptions that viewers can access with a single tap during brief pauses in playback.
}

\textbf{Description Generation.} 
The system uses Gemini-2.5 \cite{gemini_video} to generate concise descriptions tailored to the surrounding dialogue. 
For each speech segment, the corresponding video clip is provided to the model with instructions to ``\textit{(i) describe the speaker's actions, body language, and facial expressions, and (ii) include visible objects referenced in the dialogue.}'' 
The descriptions are kept brief so they can be presented with minimal interruption to playback.

\textbf{On-Demand Access.} 
During playback, a single-finger tap triggers the current description, 
which is spatialized to the speaker's location. 
Each character's descriptions are rendered using distinct voice tones to aid speaker differentiation. 
When the description ends, the video automatically resumes to minimize disruption.

\textbf{Scene Preview.} 
At the start of each video, the system provides a concise preview of the {setting} and {characters}. The setting is derived from a master shot, processed by Gemini-2.5 \cite{comanici2025gemini} to generate a second-person narration (\eg ``\textit{You are in a living room}''). Characters are introduced with brief notes on their initial actions, and their audio is spatialized to match their starting positions.

\subsection{Implementation}
\hspace*{0.8em}
\textbf{Pipeline.} 
\final{
The video-processing pipeline was implemented in Python 3.12 and deployed on a server equipped with an Intel Xeon E5-2650 CPU and an NVIDIA TITAN RTX GPU. Across the video dataset in Section~\ref{sec:technical_performance}, peak GPU memory usage was 10.5~GB, and the average processing time was 8.2 minutes per minute of input video.
}

\textbf{User Application.} 
The user application is developed in Unity 2022.3 and deployed on an iPhone 15. 
Spatial audio is rendered using Unity's Steam Audio Plugin \cite{steam_audio}, which has a built-in head-related transfer function \cite{xie2013hrtf} to produce realistic 3D sound. 
The spatialized dialogue and diegetic sounds are overlaid onto the original soundtrack to preserve the original background music. 
Textual descriptions are converted into speech using the {Volcengine Text-to-Speech Service} \cite{VolcengineTTS}, with a speech rate of 1.3. 
Additionally, the system supports two standard playback controls: 
(1) \textit{Play/Pause}: a one-finger double tap toggles video playback; and 
(2) \textit{Timeline Navigation}: a one-finger swipe up or down navigates between speech segments. 
These interactions follow screen reader conventions \cite{xu2025branch}.

\section{\textbf{Technical Evaluation}}\label{sec:technical_performance}
We evaluated \toole's pipeline in two key aspects: \textit{character trajectory accuracy} and \textit{visual description accuracy}.

\subsection{\textbf{Video Dataset}}
\begin{figure}[!h]
    \centering
    \includegraphics[width=1.0\linewidth]{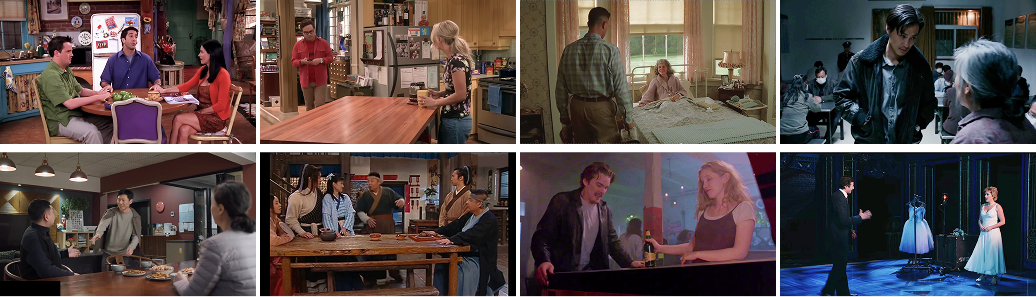}
    \caption{Examples of videos used in the technical evaluation.}
    \label{fig:4_videos}
    \Description{This figure presents examples of dialogue videos used in the pipeline evaluation. The screenshots depict diverse scenes, including casual conversations in domestic settings, interviews, classroom discussions, and theatrical performances. These varied scenarios were selected to test the system’s ability to handle different environments, character interactions, and visual contexts.}
\end{figure}

We randomly sampled 16 dialogue video clips from movies and TV shows, each recorded within a single physical space. 
The clips cover five categories: TV drama, TV comedy, movie drama, musical theater, and talk show. 
They exhibit diverse shot compositions (\eg close-ups to extreme long shots), character movements (\eg rapid walking), and lighting conditions (\eg low-light scenes). 
Each clip is one to two minutes in duration, includes two to six speakers, and contains non-overlapping speech. Figure~\ref{fig:4_videos} shows examples of the clips, and screenshots of all clips are provided in Figure~\ref{fig:video_screenshots}.

\subsection{\textbf{Character Trajectory Accuracy}}
We evaluated trajectory accuracy by labeling whether the generated trajectories matched characters' positions and movements in the original video. Each video was divided into two-second slices, and 20 slices were randomly sampled per video. A slice was marked inaccurate if any visible character's position or movement did not match the original video. One researcher labeled the data, and a second researcher reviewed the labels to ensure reliability. Accuracy was computed as the proportion of correct labels.

\begin{table}[!b]
\centering
\caption{Character trajectory accuracy across video categories, measured as the proportion of correct labels over all samples.}
\renewcommand{\arraystretch}{1.2}
\resizebox{1.0\columnwidth}{!}{%
\begin{tabular}{c|c|c}
\toprule
\textbf{Overall}: 91.9\% & \textbf{TV Comedy}: 96.2\% & \textbf{TV Drama}: 92.5\% \\
\midrule
\textbf{Movie Drama}: 86.3\% & \textbf{Musical Theater}: 87.5\% & \textbf{Talk Show}: 97.5\%\\
\bottomrule
\end{tabular}%
}
\Description{This table reports trajectory accuracy across video categories, measured as the proportion of correct labels. The first row contains Overall, TV Comedy, and TV Drama; the second row contains Movie Drama, Musical Theater, and Talk Show.}
\label{tab:trajectory_accuracy}
\end{table}

Table~\ref{tab:trajectory_accuracy} shows trajectory accuracy by category. The overall accuracy was 91.9\%. The pipeline performed well in scenes with distinct background features, enabling robust character tracking even under fast motion or sparse views. Errors mainly stemmed from two sources: (1) scarcity of full/medium shots (\eg movies dominated by close-up shots had the lowest accuracy of 86.3\%) and (2) limited feature points for multi-view alignment in dark or blurred backgrounds (\eg musical theater on a dark stage). Importantly, most inaccuracies were minor positional shifts below human auditory resolution \cite{cho2024auptimize}, and thus did not affect spatial comprehension for BLV viewers, as confirmed in our user evaluation.

\subsection{\textbf{Visual Description Accuracy}}
We evaluated description accuracy on 300 samples, consisting of all 173 descriptions from the six videos (V1–V6) used in the user study and 127 items randomly sampled from the other ten clips. A description was labeled inaccurate if it did not match the video. 
One researcher performed the initial labeling, and a second researcher reviewed the labels to ensure reliability. Accuracy was measured as the proportion of accurate labels, yielding a rate of 94.3\%.
This aligns with prior findings that shorter descriptions are less prone to hallucinations \cite{van2024making,xu2025branch}. Furthermore, because the description generation model \cite{comanici2025gemini} directly processes video clips, it effectively captures rapid gestures (\eg waving hands) and transient facial expressions. 
Occasional errors were primarily due to hallucinations \cite{favero2024multi}, which we expect to diminish with model improvements.

\section{User Evaluation}
We conducted a within-subject study with 12 BLV viewers to evaluate the effectiveness of \tool compared to a baseline. Specifically, we aimed to address the following research questions:

\begin{enumerate}[leftmargin=*, labelindent=0pt, itemindent=0pt, label=(RQ\arabic*)]
\item \textbf{Video Comprehension}: How does \tool affect BLV viewers' \textit{understanding} of dialogue videos?
\item \textbf{Viewing Experience}: How effectively does the system provide a \textit{smooth} and \textit{immersive} viewing experience?
\end{enumerate}

\begin{table*}[!h]
\centering
\caption{\rev{Both systems provide five types of information. This table presents an example recall question for each type and the corresponding functionalities provided by each system.}}
\label{tab:system_comparison_comprehension_question}
\renewcommand{\arraystretch}{1.3}
\resizebox{\textwidth}{!}{
\setlength{\tabcolsep}{3mm}
\newcommand{\hlineblack}{\specialrule{0.1em}{0em}{0em}}
\begin{tabular}{c | c | c | c}
\hlineblack
\multirow{2}{*}{\textbf{Information}} 
& \multirow{2}{*}{\textbf{Example Recall Question}}
& \multicolumn{2}{c}{\textbf{System Functionality}} \\
\cline{3-4}
& & \textbf{Sonic Stage} & \textbf{Baseline} \\
\hlineblack
Dialogue Content &
The woman asked where the man was. &
Spatialized Dialogue &
Original Dialogue
\\
\hline
Character Position &
The man was always on the right side in the scene. &
Spatialized Dialogue &
Touch-based Spatial Exploration
\\
\hline
Character Movement &
The woman moved from left to right in the scene. &
Spatialized Dialogue &
Touch-based Spatial Exploration
\\
\hline
Character Action &
The man flipped a book while speaking. &
Diegetic Sound &
Frame and Object Descriptions 
\\
\hline
Visual Detail &
The woman kept her head up all the time. &
Interactive Descriptions &
Frame and Object Descriptions 
\\
\hlineblack
\end{tabular}
}
\end{table*}

\subsection{Participants and Materials}
\subsubsection{\textbf{Participants}}
We recruited 12 BLV viewers (P1–P12; six male, six female) who regularly watched videos. 
They were recruited through a social media platform, with ages ranging from 23 to 41 years (mean = 31.9, SD = 5.3). 
Seven participants were totally blind and five were legally blind with light perception. 
All participants had normal hearing and were native Mandarin speakers.

\subsubsection{\textbf{Baseline}}
\rev{
Among existing systems for accessible video exploration (Section~\ref{sec:rw_systems}), SPICA \cite{ning2024spica} is the most comparable to \toole, as it supports both spatial and semantic access to video content. We therefore implemented a baseline modeled after SPICA. 
The baseline allows viewers to pause a video, access frame-level descriptions, and explore objects in a paused frame via touch (see Figure~\ref{fig:baseline}). 
Table~\ref{tab:system_comparison_comprehension_question} summarizes the functional differences between the two systems. 
There are two key distinctions. 
First, the baseline enables touch-based spatial exploration of paused frames in the \textit{screen space}, 
whereas \tool continuously conveys spatial information through spatial audio in the \textit{3D scene space}. 
Second, the baseline provides descriptions for all detected objects, while \tool prioritizes dialogue-relevant visual content (\ie current speakers and referenced objects). 
Both systems ran on an iPhone~15 with audio output via AirPods Pro. 
We did not compare \tool with AD, as AD does not function during dialogue, and \tool is designed to complement rather than replace AD.
}

\subsubsection{\textbf{Videos}}
To ensure a fair comparison, we selected six videos (V1–V6) from our 16-video dataset and divided them into two groups matched by genre, duration, and speech ratio. Each group contained three videos: a TV drama, a TV comedy, and a movie drama. 
The videos ranged from 1.5 to 2 minutes in length, with dialogue accounting for 87\%–97\% of the runtime. Due to this high speech density, none of the videos included AD. All videos were presented in participants’ native language (\ie Mandarin).

\subsubsection{\textbf{\final{Information Recall Test}}} 
\final{To assess participants' ability to access and retain information after interacting with each system,we developed ten recall questions for each video.} 
The questions covered five categories: 
(1) dialogue content, 
(2) character position, 
(3) character movement, 
(4) character action, and 
(5) visual detail. 
Each category included two questions per video. 
These categories were derived from key information types in AD guidelines \cite{netflix2024ADStyleGuide,videoa11y,Ofcom2024AccessServicesGuidelines}. 
\final{The questions assessed participants' immediate recall of information rather than higher-level narrative comprehension.} 
All questions were answerable using information provided by either system. 
Table~\ref{tab:system_comparison_comprehension_question} provides example questions. 
Participants answered by selecting one of three options: ``\textit{yes}'', ``\textit{no}'', or ``\textit{I don't know}'', with ``\textit{I don't know}'' counted as incorrect.

\subsection{Design and Procedure}
\subsubsection{\textbf{Procedure}}
Participant first provided demographics and completed a 10-minute tutorial on both systems. 
They then proceeded to the main study. Each participant used both systems to view dialogue videos. 
To ensure study control, 
the four combinations of video groups and systems were evenly distributed across the twelve participants, with videos presented in random order. 
Participants watched videos as they would in daily life, freely using features and replaying content as needed, and answered ten recall questions after each video without rewatching. 
Upon completing both systems, participants filled out a questionnaire (Figure~\ref{fig:comparison_ratings}) and provided qualitative feedback in a semi-structured interview. 
Each session lasted about 90 minutes, was conducted one-on-one in person, and participants received compensation of approximately 15 USD in local currency for their time.

\subsubsection{\textbf{Metrics}}
We used both objective and subjective measures to evaluate the systems. Objective video comprehension was quantified by accuracy on recall questions, analyzed by category. 
Viewing experience was assessed using established scales for narrative engagement \cite{busselle2009measuring} and spatial presence \cite{hartmann2015spatial}. 
Usability was measured with relevant items from the system usability scale \cite{lewis2018system}. All subjective measures used a seven-point Likert scale.

\subsubsection{\textbf{Analysis}}
We collected audio recordings, interaction logs, and questionnaire data during the study. 
Quantitative data that satisfied the assumptions of normality and equal variance were analyzed using paired t-tests. 
Subjective ratings were assessed using the Wilcoxon signed-rank test \cite{wilcoxon1970critical}. 
Audio recordings were transcribed verbatim and analyzed using reflexive thematic analysis \cite{braun2019reflecting} to address the two research questions (RQ1–RQ2).

\section{User Evaluation Results}
We conducted  a total of 72 trials (12 users $\times$ 3 videos $\times$ 2 systems). 
For each system, participants answered 360 factual recall questions, yielding 72 responses per category. 
The following sections present results for video comprehension and viewing experience.

\subsection{Video Comprehension (RQ1)}

\begin{figure*}[h]
    \centering
    \includegraphics[width=0.95\linewidth]{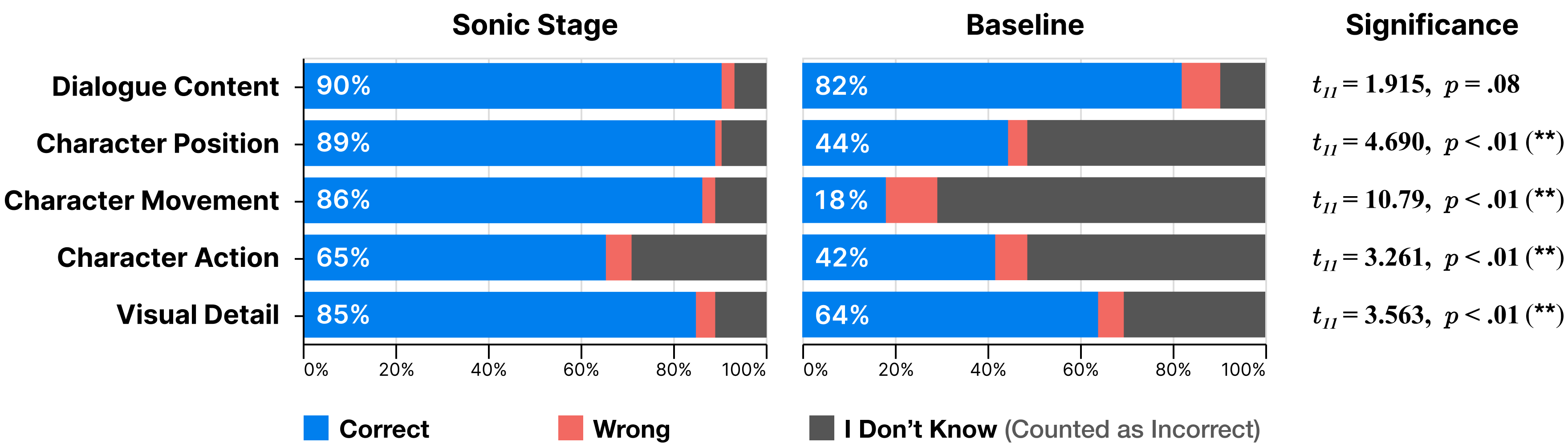}
    \caption{Accuracy rates on recall questions by category. Each category included 72 responses per system. 
    Statistical significance was assessed using paired t-tests.}
    \label{fig:comprehension_results}
    \Description{This figure compares comprehension accuracy rates across five categories—Dialogue Content, Character Position, Character Movement, Character Action, and Visual Detail—for Sonic Stage and a baseline system. Each category includes 72 responses. Results show that Sonic Stage substantially improves both spatial understanding (\eg positions and movements) and semantic understanding (\eg actions and visual details). Statistical significance, assessed with paired t-tests, indicates highly significant improvements in four categories (p < .01).}
\end{figure*}

\begin{figure*}[!th]
    \centering
    \includegraphics[width=0.95\textwidth]{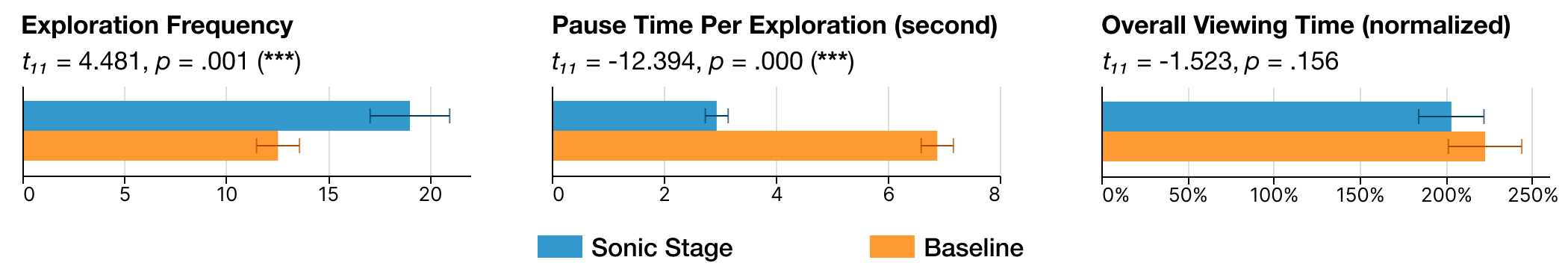}
    \caption{Usage patterns for each system. \tool led to more frequent exploration with shorter pauses, without increasing total viewing time. Statistical significance was evaluated using paired t-tests. Error bars represent standard errors.}
    \label{fig:usage_data}
    \Description{This figure shows usage patterns comparing Sonic Stage and the baseline system. Sonic Stage resulted in significantly higher exploration frequency and shorter pause times per exploration, while overall viewing time remained comparable between systems. Statistical significance was assessed using paired t-tests, with error bars representing standard errors.}
\end{figure*}

\textbf{\tool significantly enhanced participants' spatial and semantic understanding of dialogue videos.} 
Figure~\ref{fig:comprehension_results} shows the accuracy rate of recall questions for each category. 
Compared to the baseline, \tool improved accuracy in four key aspects: 
character position ($t_{11} = 4.690, \ p < .01$), 
movement ($t_{11} = 10.79, \ p < .01$),
action ($t_{11} = 3.261, \ p < .01$), 
and visual details ($t_{11} = 3.563, \ p < .01$). 
Subjective ratings (Figure~\ref{fig:comparison_ratings}, Q2 -- Q5) echoed these results, indicating perceived improvements in the same categories. 
No significant difference was found in dialogue comprehension accuracy ($t_{11} = 1.915,\ p = .08$), 
although participants reported that dialogue was easier to follow with \tool ($Z = -2.428,\ p < .05$), because ``\textit{the descriptions did not disrupt the dialogue flow}'' (P1). 
Overall, participants indicated that \tool conveyed essential information more effectively than the baseline ($Z = -3.108, \ p < .01$). 
In the following, we provide a detailed analysis of participants' spatial and semantic understanding.

\subsubsection{\textbf{Spatial Understanding}}\label{sec:spatial_understanding}
With \toole, participants achieved an accuracy of 89\% for position and 86\% for movement. 
Most errors were ``\textit{I don’t know}'' responses, typically attributed to forgetting related information. 
In contrast, the baseline condition resulted in significantly lower accuracy: 44\% for position and 18\% for movement. Most baseline errors were also ``\textit{I don't know}'' responses, which participants attributed to confusing camera changes (twelve mentions) and long pauses for spatial exploration (ten mentions).

\textbf{\tool supported intuitive spatial understanding by maintaining a \textit{coherent} spatial representation.} 
Participants reported that scene-anchored spatial audio made it ``\textit{much easier to picture how characters move around in the scene}'' (P5). 
Because spatial relationships remained stable over time, \tool ``\textit{reduced the load to piece together character positions from separate frames}'' (P4), 
enabling participants to ``\textit{form a vivid spatial mental map}'' (P9). 
In contrast, although the baseline enabled spatial exploration within paused key frames, participants described the process as time-consuming: ``\textit{I need to swipe across the screen to see if I missed someone}'' (P1). 
Moreover, because spatial information was presented in screen space, camera changes often caused confusion. 
As P3 noted: ``\textit{In the previous frame, there were two persons covering small portions, but now it was one person covering almost the entire frame. I got really confused. Did anyone move?}'' 
Such inconsistency made it ``\textit{nearly impossible to infer character movements}'' (P1).

\subsubsection{\textbf{Semantic Understanding}}\label{sec:semantic_understanding}

When using \toole, participants achieved 65\% accuracy on character actions and 85\% on visual details, compared to baseline scores of 42\% and 64\% respectively. 
Across both systems, the most frequent semantic understanding error was the response ``\textit{I don't know}'' (see Figure~\ref{fig:comprehension_results}), which typically arose when the relevant information was accessible only through on-demand descriptions. 
Participants performed better on questions about visual details than character actions, likely because visual details (\eg clothing) persist across descriptions, while brief actions (\eg flipping a book) only occur at specific moments. 

\textbf{\tool supported viewers' contextual needs through \textit{dialogue-relevant} descriptions.} 
Participants valued that \tool prioritized descriptions based on the audio track, making it ``\textit{quite easy to access the most relevant information}'' (P2). 
As P9 explained, ``\textit{Because I get most information from the audio, what I care about most is also centered on it --- like what the speaker is like and what they're talking about. That information is right there when I need it.}'' 
In contrast, while the baseline provided detailed frame and object descriptions, participants felt it lacked prioritization. As P5 noted, ``\textit{It tries to describe everything, but but when everything is highlighted, nothing really stands out.}'' 

\subsubsection{\textbf{Exploration Pattern}}\label{sec:exploration_pattern}
We analyzed how participants initiated exploration, defined as triggering descriptions in \tool or examining key frames in the baseline, with continuous pauses counted as a single instance. 
Figure~\ref{fig:usage_data} presents usage patterns for each system. 
Participants initiated exploration significantly more often with \tool (M = 19.0, SD = 6.8) than with the baseline (M = 12.6, SD = 3.7; $t_{11} = 4.48,\ p = .001$). 
Pauses per exploration were also shorter with \tool (M = 2.95s, SD = 0.71) compared to the baseline (M = 6.88s, SD = 1.00; $t_{11} = -12.39,\ p < .001$). 
Notably, the overall viewing time, normalized by each video's duration, did not differ significantly between systems ($t_{11} = -1.52,\ p = .156$). 
This suggests that participants explored more frequently with \toole, with shorter pauses during each exploration. 

\textbf{\tool encouraged exploration through \textit{audio-guided} engagement.} 
Participants noted that salient auditory cues (\eg moving speech or diegetic sounds) ``\textit{sparked their interest in accessing details}'' (P5). 
These audio cues made them ``\textit{more aware of when to explore}'' (P7), thereby ``\textit{reducing fear of missing interesting moments}'' (P11) and ``\textit{lowering decision-making efforts}'' (P6).
Furthermore, participants reported that \tool made exploration feel ``\textit{less intrusive and more enjoyable}'' (P9), because the descriptions were ``\textit{short, concise, and directly relevant to the audio}'' (P1). 
In contrast, they found examining objects in the baseline time-consuming, making them ``\textit{less willing to explore}'' (P3).

\subsection{Video Viewing Experience (RQ2)}

\textbf{\tool achieved high usability ratings.} 
All participants rated \tool at the maximum score for both ease of use and ease of learning (see Figure~\ref{fig:quality_rating}). Compared to the baseline, they found \tool significantly easier to use ($Z = -2.032$, $p < .05$), requiring less mental effort ($Z = -2.877$, $p < .01$) and less physical effort ($Z = -2.889$, $p < .01$). As P3 noted, ``\textit{Most of the information is well integrated into the audio, and getting additional details is as simple as a tap.}'' Consequently, they expressed a strong willingness to use the system in the future ($\mu = 6.91$ on a seven-point scale).

\textbf{Participants perceived the spatial audio as distinguishable and smooth.} 
Figure~\ref{fig:quality_rating} shows participant ratings of the audio quality. 
For spatial audio, they rated distinguishability at 6.25 (SD = 0.75) and smoothness at 6.42 (SD = 0.67) on a seven-point scale, indicating high spatial audio quality. 
For diegetic sound, participants rated coherence at 5.92 (SD = 0.79) and clarity at 5.00 (SD = 1.28). 
The relatively lower clarity score was due to the ambiguity of certain diegetic sounds. As P3 noted, ``\textit{Some sounds, like plastic wrap, do not clearly convey what is happening. In such cases, the interactive description is important for clarifying the associated actions.}''

\textbf{\tool enhanced BLV viewers' spatial presence in the virtual scene.} 
Participants reported feeling more present within the scene ($Z = -3.090,\ p < .01$) and experiencing a stronger sense of being surrounded by characters ($Z = -3.072,\ p < .01$). 
All participants highlighted that spatial audio made them feel ``\textit{on the stage}'' (P1) and ``\textit{in the scene}'' (P3). As P10 described, ``\textit{It truly feels like I'm sitting at the table and they are conversing around me.}'' 

\textbf{Participants reported higher narrative engagement with \toole.} 
They perceived a richer story world ($Z = -3.084,\ p < .01$) and reported deeper narrative immersion ($Z = -3.020,\ p < .01$). 
Participants emphasized how the spatial soundtrack heightened emotional resonance. As P1 noted, ``\textit{The spatial sound has a stronger emotional impact. It brings out the characters' emotions and makes their conflict feel more real.}'' 
Similarly, P9 noted, ``\textit{When the speaker approaches me as she calls out, it feels like a puppy leaping forward. I can strongly feel her excitement in that moment.}''

\textbf{Consequently, \tool delivered a more enjoyable viewing experience.} 
Participants rated the overall viewing experience with \tool as significantly smoother ($Z = -3.097,\ p < .01$) and more enjoyable ($Z = -2.716,\ p < .01$) compared to the baseline. 
As P5 noted, ``\textit{I really enjoyed how the dialogue, the sounds, and the descriptions worked together to create a seamless experience. It felt like a stage production I would really want to attend.}'' 
\section{Discussion}
Audio description (AD) is constrained not to overlap with speech, which often prevents BLV viewers from accessing essential visual information during dialogue. 
To address this challenge, \tool introduces three auditory techniques: 
\textit{spatialized dialogue}, \textit{diegetic sound}, and \textit{interactive descriptions}. 
User evaluations showed that \tool significantly improved video comprehension, spatial presence, and narrative engagement. 
In this section, we discuss five key aspects of \toole: 
(1) the balance between user agency and narrative coherence; 
(2) the effects of its design components; 
(3) its integration with AD;
(4) the design of immersive audio representations; and
(5) its potential extension to other video genres.

\subsection{\final{Balancing Agency and Narrative Coherence}}
\final{
Accessible video exploration often involves balancing user agency with narrative coherence \cite{ning2024spica,cheema2025describe,xu2025branch,xu2025danmua11y}.
\tool addresses this tension by enabling BLV viewers to receive essential information during playback while retaining the option to pause briefly for additional detail. 
Building on our findings and prior work \cite{jiang2023_beyond_AD,xu2025branch,amy2020rescribe}, we identify two directions for further improving this balance.
}

\final{
\textbf{Avoiding excessive auditory cues.} 
Across our 16-video dataset, Sonic Stage generated 41 diegetic sounds, corresponding to an average of 1.8 sounds per minute. 
The shortest interval between generated sounds was 6.5 seconds, and no participants reported listening discomfort. 
Yet, when several actions occur in rapid succession or simultaneously, auditory cues may become difficult to distinguish. 
Future work should therefore explore methods for prioritizing the most salient actions and provide customization options that allow users to adjust cue frequency.
}

\final{
\textbf{Supporting flexible interaction styles.} 
In our study, ten participants valued on-demand exploration, whereas two preferred experiences that required less active interaction.
Future systems could therefore support multiple interaction styles:
(a) \textit{user-driven approaches}, in which users decide when to request additional information;
(b) \textit{mixed-initiative approaches}, in which the system suggests potential moments for exploration \cite{xu2025branch}; and 
(c) \textit{system-driven approaches}, in which the system automatically extends the video timeline to present additional descriptions \cite{amy2020rescribe}. 
Offering these alternatives would allow users to select an interaction style that better matches their viewing needs and preferences \cite{jiangchi24context}.
}

\subsection{\final{Effects of \toole's Design Components}}
\final{
In our evaluation, \tool differed from the baseline in three key respects. First, \tool used \textbf{scene-space spatialization} rather than screen-space spatialization. 
Second, it incorporated \textbf{inline auditory cues}, including spatialized dialogue and diegetic sounds, rather than relying solely on on-demand descriptions. 
Third, it provided \textbf{dialogue-relevant visual descriptions} rather than attempting to describe every visible object.
}

\final{
Participants’ feedback indicated how each of these design choices shaped their experiences. First, scene-space spatialization maintained spatial audio continuity across camera cuts, supporting a more coherent understanding of characters' positions and movements. 
Second, inline auditory cues helped participants identify when additional visual information was available and decide when to explore it, thereby supporting audio-guided engagement. 
Third, dialogue-relevant visual descriptions surfaced information that participants considered important while minimizing disruptions to the flow of dialogue. 
Together, these findings suggest that, relative to the baseline, \toole’s design choices supported more coherent spatial understanding, audio-guided exploration, and a more continuous viewing experience. 
Yet, our study did not isolate the effects of individual design components; future work should evaluate their respective contributions through component-level studies.
}

\subsection{Integrating \tool with AD}

\begin{figure}[h]
    \centering
    \includegraphics[width=0.9\columnwidth]{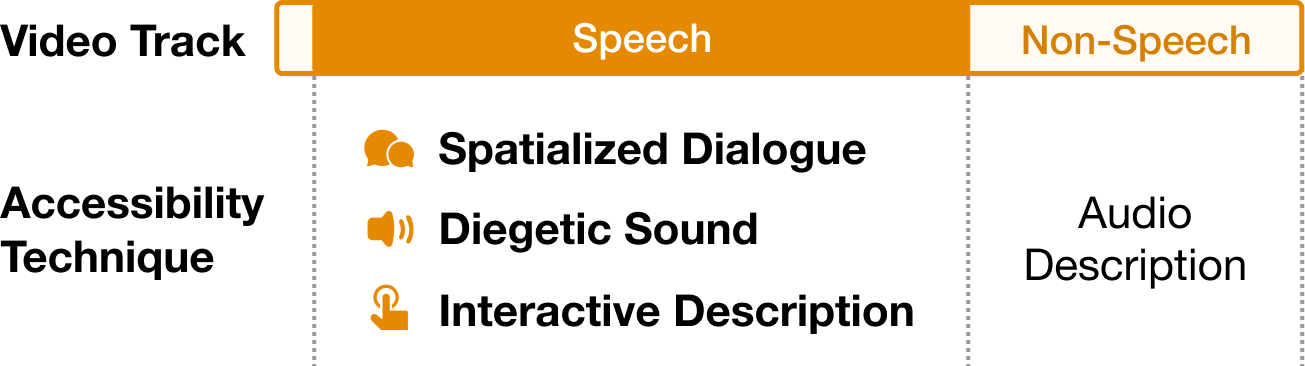}
    \caption{\rev{The approach for integrating \tool with AD.}}
    \label{fig:integration}
\end{figure}

\rev{
When integrating \tool with AD, \toole's three auditory techniques convey visual information during speech, while AD operates during non-speech segments (see Figure~\ref{fig:integration}). 
This integration allows each technique to leverage its strengths without competing for temporal space. 
For instance, spatialized dialogue communicates spatial layout information in real time, enabling AD to focus on other salient visual elements. 
Furthermore, effective integration requires minimizing redundancy across modalities \cite{xu2025branch}. 
Future work may explore methods that prevent interactive descriptions from repeating information already conveyed through AD. 
Beyond coexisting with AD, the three auditory techniques could also be leveraged to enhance AD itself. 
For instance, spatializing AD based on referenced object locations could further improve spatial awareness for BLV viewers \cite{chang2022omniscribe}.
}

\subsection{Toward Immersive Audio Representations}\label{sec:improvements_spatial} 
Our study identifies several directions for advancing the immersive audio representations of \toole. 
First, future systems could \textbf{support alternative listening perspectives}. 
While \tool currently positions listeners near the stage center to evoke the sensation of ``\textit{being on the stage}'' (P1), offering other placements (e.g., an audience perspective) or allowing movement within the soundscape could broaden experiential possibilities for BLV viewers. 
A second direction is to \textbf{enrich scene representations}. While the system currently sonifies active speakers, participants also expressed interest in perceiving other characters and objects. Future work could introduce additional cues, such as footsteps indicating the presence of non-speaking characters \cite{jain2023frontrow}. 
A third direction is to \textbf{amplify subtle movements}. During our design process, BLV viewers expressed interest in perceiving subtle motions (e.g., swaying) via spatial audio. 
However, after testing amplified motion cues, we chose not to adopt this approach because it produced misleading interpretations: ``\textit{It made me think the person had moved drastically}'' (SD2). 
Future systems may draw on film sound theories \cite{balazs1985theory,chion2019audio} to balance distinction with accuracy. 
Participants also emphasized the need to \textbf{enhance front–back distinction}, which is challenging due to inherent limitations of human auditory systems \cite{cho2024auptimize}. 
Additional cues, such as pitch variation \cite{chheda2025brought}, may help resolve front-back ambiguities. 
Finally, while \tool currently supports videos recorded within the same scene, future designs should \textbf{support transitions across scenes}. For example, ambient sound effects could be used to signal scene changes \cite{donker2002design,lopez2021enhancing}, enabling BLV users to perceive transitions smoothly.

\subsection{Extending \tool to Diverse Genres}\label{sec:future_work}

\begin{figure}[!h]
    \centering
    \includegraphics[width=1.0\columnwidth]{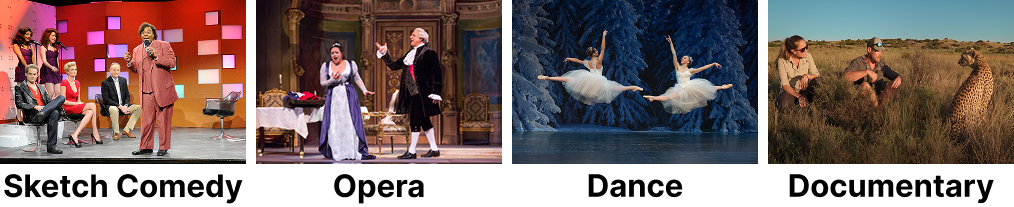}
    \caption{\tool can be extended to diverse video genres.}
    \Description{This figure presents examples of video genres that BLV viewers envision applying Sonic Stage to, including opera, concert, sketch comedy, dance performance, and documentary. These genres highlight the broad potential of the system across diverse video genres.}
    \label{fig:extended_app}
\end{figure}

BLV viewers in our study identified several contexts where \tool could enhance their viewing experiences (see Figure~\ref{fig:extended_app}). 
The first was \textit{stage performances}, 
such as sketch comedy or opera. 
Participants hoped that \tool could help them ``\textit{perceive performers' movements directly}'' (P3). 
Another opportunity is being able to appreciate \textit{dance and gymnastics}. 
Participants envisioned that \tool could make dancers' gestures and movements accessible by translating them into spatial sound. The key challenge, however, is to sonify complex body movements without overwhelming viewers. 
The third opportunity is using \tool for more \textit{immersive documentaries} and virtual travel experiences; that is, to ``\textit{experience a novel world}'' (P6). These genres often depict more open-ended scenes, introducing new challenges for designing spatial audio cues. 
In the following, we outline open challenges for future research.

\textbf{The first challenge is designing accessible audio representations for open-ended scenes.} 
Currently, \tool generates soundscapes anchored to fixed physical spaces. 
Yet, this approach that does not generalize well to open-ended scenes. 
For instance, when two characters converse while walking in a desert, 
spatialization may emphasize their relative motion rather than rely on fixed scene anchors. 
From a technical perspective, open-ended scenes often lack visual anchors for 3D reconstruction. 
One potential solution is to generate \textit{spatial mental maps} that reflect how sighted viewers perceive spatial information \cite{yin2025spatial, chen2024spatialvlm}.

\textbf{Another challenge lies in sonifying complex body movements}. 
In dance performances, spatial audio can represent a dancer's location on stage but cannot capture the movements of individual body parts. Future work could how to sonify complex body movements to BLV viewers \cite{chen2025voice,jiang2025audio}. 
Advances in reconstructing body meshes \cite{delmas2022posescript} may also support more precise movement descriptions. 
Furthermore, when multiple dancers perform simultaneously, auditory cues must be carefully designed to avoid clutter \cite{guerreiro2023_audio_design_blind}. To address this, future research could investigate interaction techniques that allow viewers to selectively focus on individual performers.

\textbf{The third challenge is separating concurrent audio sources}. 
Currently, \tool assumes non-overlapping speech. Future work should address more demanding conditions, such as concurrent voices or uneven volumes \cite{chen2025rhythmta}. 
While traditional audio processing methods often degrade audio quality when separating overlapping speech, voice cloning models \cite{arik2018neural} offer a promising direction for reconstructing high-quality individual voices.

\subsection{Limitations and Future Work}
\tool focuses on conveying visual information through audio. Future work could investigate complementary feedback modalities, such as haptics \cite{butler2021technology,xu2020virtual}. 
\final{\tool currently does not communicate camera cuts to BLV users; future research could explore how such transitions can be conveyed without increasing cognitive load.} 
Our study primarily examined short video clips; applying \tool to longer-form content may introduce new challenges, such as viewer fatigue \cite{wang2026wearable}, which warrants further investigation. 
\final{While our evaluation measured participants’ immediate recall of factual information, future studies could examine how \tool influences higher-level narrative comprehension, such as understanding characters' relationships.} 
Although this work emphasizes automatic approaches for generating accessible audio cues, future research could also support video creators in authoring accessible audio cues during production (e.g., \cite{mimosa_spatial_audio}). 
While \tool leverages spatial audio, it may inadvertently reduce accessibility for viewers who are hard of hearing. Additionally, the use of scene-space spatial audio could confuse sighted or partially sighted viewers; for example, changes in camera angle may introduce mismatches between visual and auditory information. 
Future systems should investigate strategies that accommodate diverse hearing and visual abilities across different user populations.
\section{Conclusion}
We have presented \toole, a system that conveys essential visual information during dialogue through three auditory techniques: \textit{spatialized dialogue}, \textit{diegetic sound}, and \textit{interactive descriptions}. 
Powered by a fully automated pipeline, these techniques complement audio description (AD) and provide an immersive video experience for BLV audiences. 
User evaluations demonstrated that \tool effectively enhanced video comprehension, spatial presence, and narrative engagement. 
Our study also identifies opportunities for extending this approach to other video genres, such as documentaries, concerts, and dance performances. 
We hope this work offers insights into the design of immersive and interactive audio representations and inspires future research that enables deeper engagement with visual content for BLV audiences.

\begin{acks}
The authors thank all participants for their support during the studies and are grateful to the reviewers for their constructive feedback.
\end{acks}

\bibliographystyle{ACM-Reference-Format}
\bibliography{references}

\appendix
\newpage
\onecolumn
\section{Appendices}

\subsection{Subjective Ratings in the Evaluation Study}
\begin{figure*}[h]
    \centering
    \includegraphics[width=\textwidth]{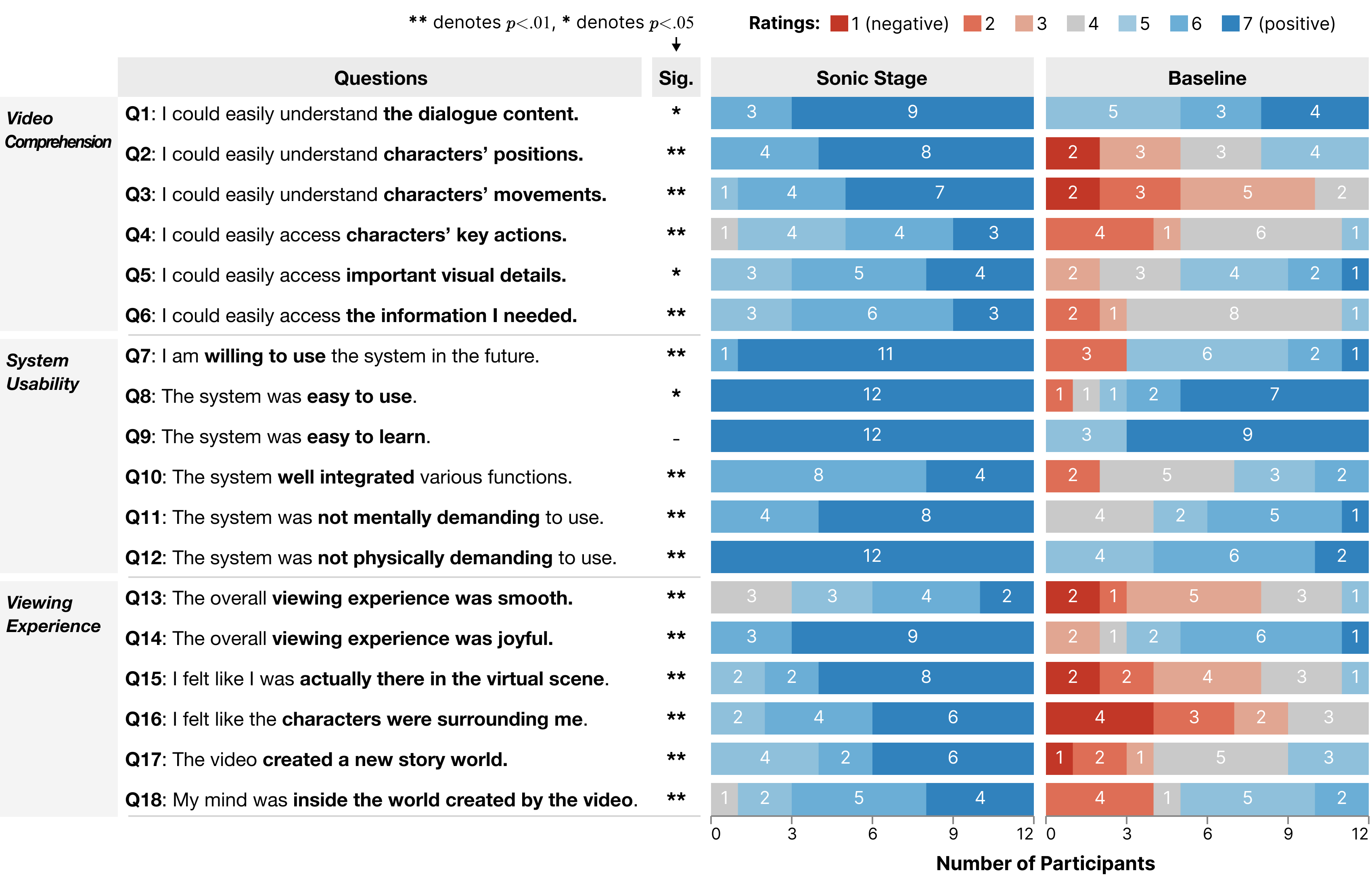}
    \caption{Participant rating distributions for both systems (1 = strongly negative, 7 = strongly positive). Asterisks indicate statistical significances based on the Wilcoxon signed-rank test.}
    \label{fig:comparison_ratings}
    \Description{This figure shows participant rating distributions for Sonic Stage and the baseline system across three dimensions: Video Comprehension (Q1–Q6), System Usability (Q7–Q12), and Viewing Experience (Q13–Q18). Ratings range from 1 (strongly negative) to 7 (strongly positive). Results indicate consistently higher ratings for Sonic Stage, with significant improvements in comprehension, usability, and immersive experience. Statistical significance was assessed using the Wilcoxon signed-rank test.}
\end{figure*}

\begin{figure*}[h]
    \centering
    \includegraphics[width=0.85\textwidth]{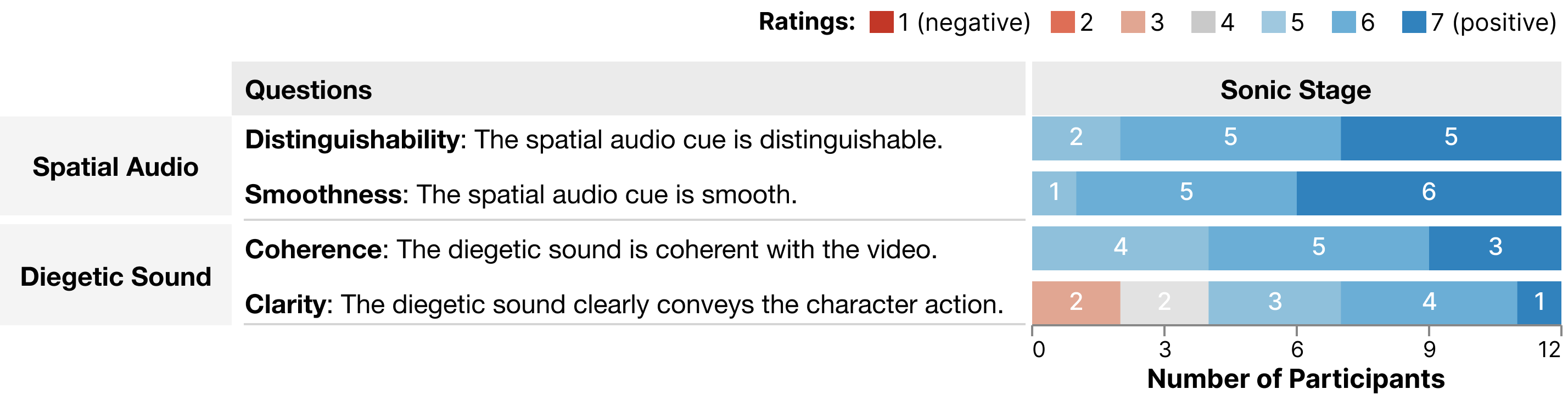}
    \caption{Participant ratings of the audio quality in \tool (1 = strongly negative, 7 = strongly positive).}
    \label{fig:quality_rating}
    \Description{This figure presents participant ratings of spatial audio and diegetic sound quality in Sonic Stage. Ratings range from 1 (strongly negative) to 7 (strongly positive). For spatial audio, participants evaluated distinguishability and smoothness, while for diegetic sound, they rated coherence with the video and clarity in conveying character actions. Overall, ratings were strongly positive, with most participants selecting values between 5 and 7.}
\end{figure*}

Figure~\ref{fig:comparison_ratings} and Figure~\ref{fig:quality_rating} provide subjective rating results from the evaluation study. 

\newpage
\subsection{Videos Used for the Evaluation}
\begin{figure}[!h]
    \centering
    \includegraphics[width=0.98\textwidth]{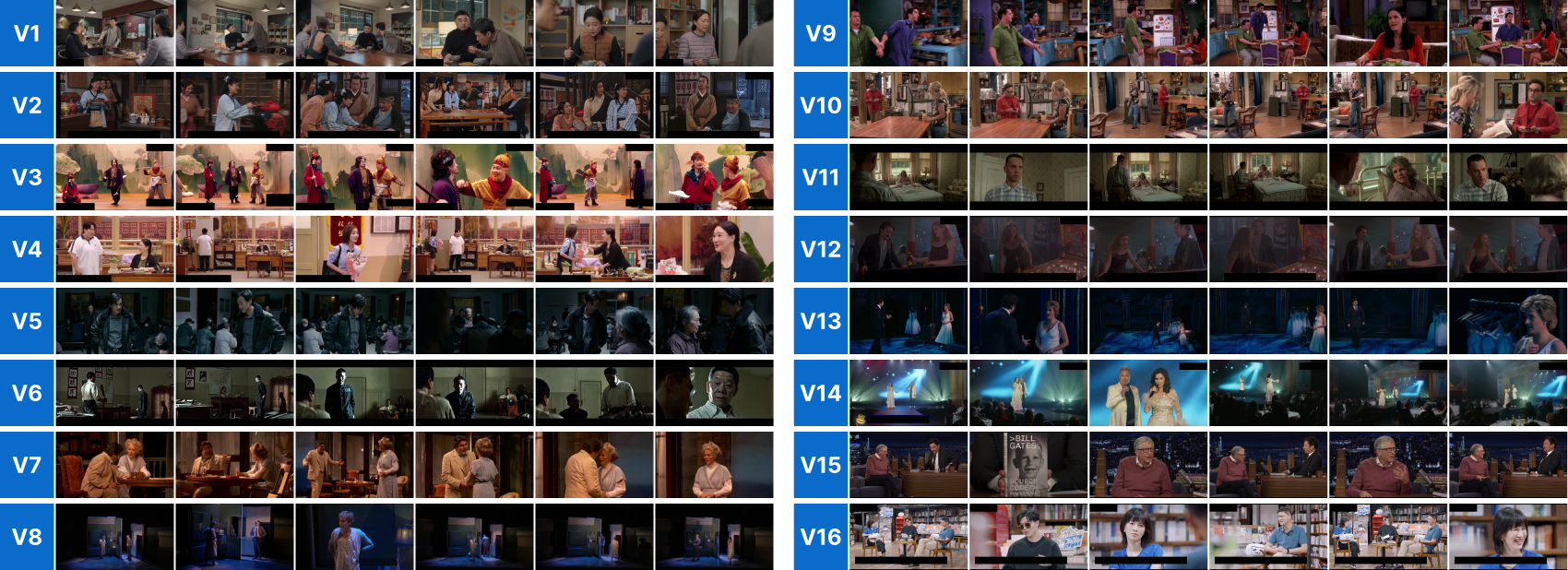}
    \caption{Screenshots from the 16-video dataset used for system evaluation.}
    \Description{This figure shows the video screenshots for the sixteen videos used in the evaluation study. Each row corresponds to one video, showing multiple frames that illustrate the diversity of settings, characters, and visual contexts represented across the dataset.}\label{fig:video_screenshots}
\end{figure}

Table~\ref{tab:video_dataset} and Figure~\ref{fig:video_screenshots} present descriptions and screenshots of the 16 videos used in the evaluation study. 

\begin{table}[htbp]
\centering
\caption{Information on the 16-video dataset used for evaluation. V1–V6 are used for user evaluation.}
\resizebox{1.0\columnwidth}{!}{
\renewcommand{\arraystretch}{1.3}

\begin{tabular}{cccccc}
\toprule
\textbf{VID} & \textbf{Type} & \textbf{Title} & \textbf{Description} & \textbf{Length} & \textbf{Dialogue Ratio} \\
\midrule
\textbf{V1} & \textbf{TV Drama}    & \textbf{Song of Ordinary People} & \textbf{Four people chatting over dinner.} & \textbf{01:41} & \textbf{89\%} \\
\textbf{V2} & \textbf{TV Drama}    & \textbf{My Own Swordsman}        & \textbf{Six people talking in a hostel.}   & \textbf{01:32} & \textbf{90\%} \\
\textbf{V3} & \textbf{TV Comedy}   & \textbf{Journey to the West}     & \textbf{Three people arguing on stage.} & \textbf{01:50} & \textbf{89\%} \\
\textbf{V4} & \textbf{TV Comedy}   & \textbf{Farewell, My Teacher}    & \textbf{Three people chatting in an office.} & \textbf{01:37} & \textbf{97\%} \\
\textbf{V5} & \textbf{Movie Drama} & \textbf{I'm Not a Medicine God}  & \textbf{Two people talking in a police station.}   & \textbf{02:14} & \textbf{87\%} \\
\textbf{V6} & \textbf{Movie Drama} & \textbf{The Last Emperor}        & \textbf{Three people discussing in a prison.} & \textbf{01:35} & \textbf{89\%} \\
V7  & TV Drama    & Long Day's Journey into Night & Two people talking in a room.  & 01:11 & 99\% \\
V8  & TV Drama    & Angels in America             & Two people arguing at home.  & 00:57 & 96\% \\
V9  & TV Comedy   & Friends                       & Three people arguing in a room. & 00:52 & 87\% \\
V10 & TV Comedy   & The Big Bang Theory           & Two people chatting at home. & 00:52 & 94\% \\
V11 & Movie Drama & Forrest Gump                  & Two people talking by a bed. & 01:44 & 88\% \\
V12 & Movie Drama & Before Sunset                 & Two people chatting in a bar.  & 01:48 & 96\% \\
V13 & Musical Theater  & Diana: the musical        & Two people arguing in a hall. & 01:00 & 97\% \\
V14 & Musical Theater  & Elisabeth: the musical    & Two people singing on stage. & 01:14 & 96\% \\
V15 & Talk Show   & The Tonight Show              & Two people talking in a studio.  & 01:11 & 95\% \\
V16 & Talk Show   & The Bilibili Interview        & Three people chatting in a studio. & 02:14 & 97\% \\
\bottomrule
\end{tabular}
}
\label{tab:video_dataset}
\end{table}

\newpage
\subsection{Participant Demographics}
\begin{table*}[!htbp]
  \centering
  \caption{Participants' demographics. Two BLV Sound Designers (SD1 and SD2) contributed to the iterative system design process. Participants P1–P12 participated in the evaluation study. 
  \textbf{Spatial audio} refers to participants' self-reported level of familiarity with spatial audio, rated on a five-point scale \ (1 = not familiar, 2 = slightly familiar, 3 = moderately familiar, 4 = familiar, and 5 = highly familiar).}
  \resizebox{0.9\columnwidth}{!}{
  \renewcommand{\arraystretch}{1.3}
  \setlength{\tabcolsep}{2mm}{
    \begin{tabular}{c c c c c l l}
      \toprule
      \textbf{PID} & \textbf{Age} & \textbf{Gender} & \textbf{Visual Condition} & \textbf{Video Viewing} & \textbf{Occupation} & \textbf{Spatial Audio}\\
      \midrule
      SD1 & 35 & M & Totally blind & Daily & Sound designer & 4: Familiar\\
      SD2 & 32 & M & Legally blind & Daily & Sound designer & 5: Highly familiar\\
      P1 & 30 & M & Legally blind & Daily & Music Performer & 4: Familiar\\
      P2 & 34 & F & Totally blind & 4–5 times per week & Music Performer & 5: Highly familiar\\
      P3 & 37 & M & Legally blind & Daily & Sound designer & 5: Highly familiar\\
      P4 & 27 & F & Totally blind & 4–5 times per week & Salesperson & 4: Familiar\\
      P5 & 31 & F & Legally blind & Daily & Piano Tuner & 4: Familiar\\
      P6 & 29 & M & Legally blind & Daily & Salesperson & 4: Familiar\\
      P7 & 24 & M & Totally blind & Daily & Video Creator & 3: Moderately familiar\\
      P8 & 39 & F & Totally blind & Daily & Homemaker & 2: Slightly familiar\\
      P9 & 35 & F & Totally blind & 2–3 times per week & Freelancer & 2: Slightly familiar\\
      P10 & 32 & M & Legally blind & Daily & Video Creator & 5: Highly familiar\\
      P11 & 23 & F & Totally blind & Daily & Student & 3: Moderately familiar\\
      P12 & 41 & M & Totally blind & 4–5 times per week & Guitarist & 4: Familiar\\
      \bottomrule
    \end{tabular}
    }
  }
  \label{tab:demographics}
\end{table*}

\subsection{Implementation Details of the Baseline System}\label{sec:baseline-detail}
\begin{figure}[!h]
    \centering
    \includegraphics[width=1.0\columnwidth]{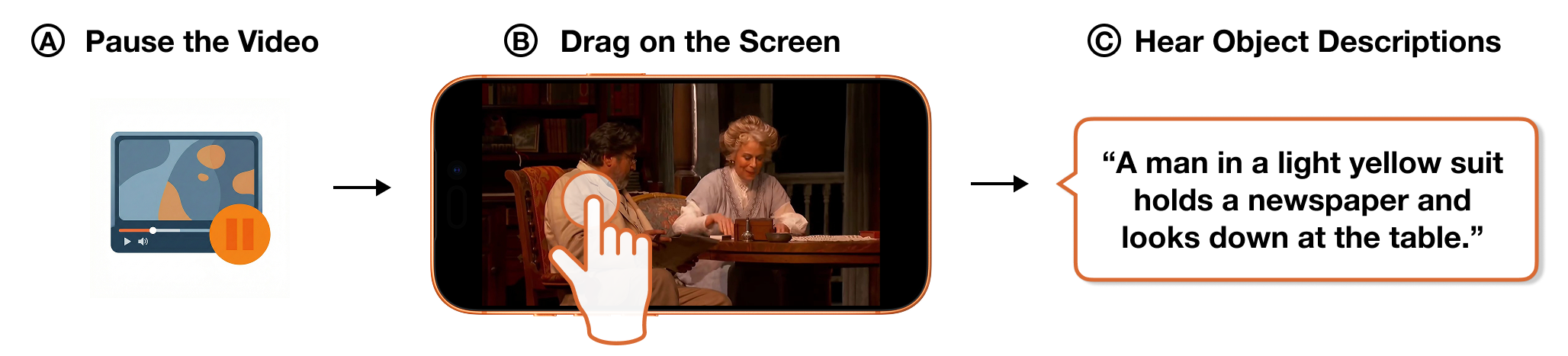}
    \caption{The baseline system supports touch-based spatial exploration. Users can (A) pause the video, (B) move their fingers across the touchscreen, and (C) hear descriptions of the objects they touch.}
    \label{fig:baseline}
\end{figure}

The baseline system is deployed on an iPhone 15 and supports the following interactions: 
(1) \textit{Play/Pause:} A one-finger double tap toggles video playback between play and pause states. 
(2) \textit{Key Frame Navigation:} A one-finger swipe upward or downward navigates between key frames, while a single tap announces the description of the current frame. 
(3) \textit{Spatial Exploration:} When playback is paused at a key frame, users can drag one finger across the screen to hear descriptions of the objects being touched (see Figure~\ref{fig:baseline}). 
All descriptions are generated using the method in the SPICA system \cite{ning2024spica} and are delivered through mono audio.

\newpage
\subsection{Technical Details of \toole}

\begin{figure*}[!h]
    \centering
    \includegraphics[width=0.95\linewidth]{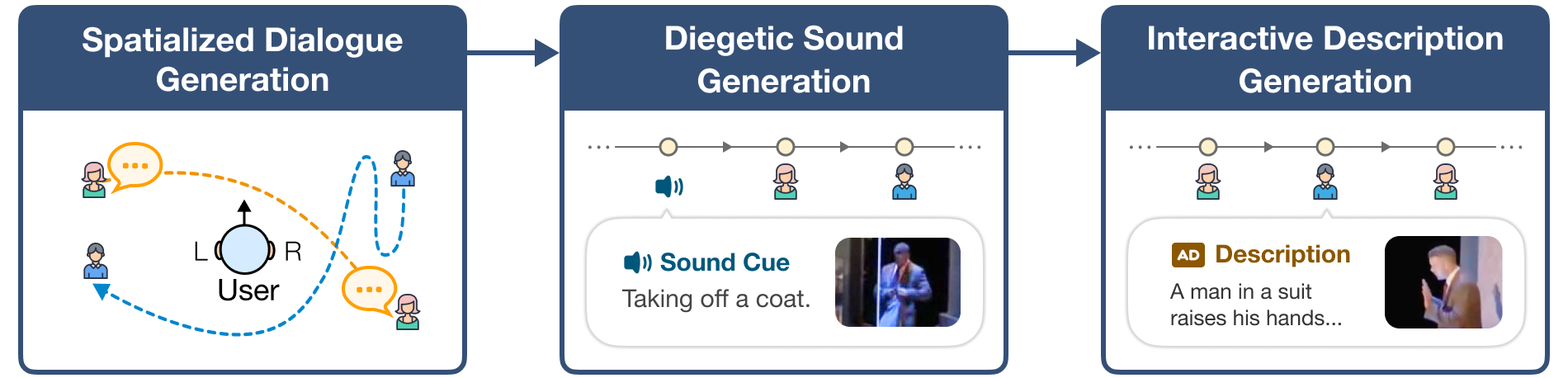}
    \caption{The \tool pipeline comprises three core modules. The first module applies 3D reconstruction to recover character trajectories and spatialize their dialogue. The second module employs text-to-sound models to generate diegetic sound for character actions. The third module uses multimodal language models to produce descriptions related to the dialogue.}
    \label{fig:pipeline}
    \Description{This figure illustrates the Sonic Stage pipeline, which generates immersive auditory experiences in three stages. The first module, spatialized dialogue generation, uses 3D reconstruction to track character positions and spatialize their speech. The second module, diegetic sound generation, employs text-to-sound models to produce sound cues for detected character actions, such as “taking off a coat.” The third module, interactive description generation, leverages multimodal language models to create contextual descriptions that are synchronized with the dialogue.}
\end{figure*}

\begin{figure*}[!h]
    \centering
    \includegraphics[width=0.95\linewidth]{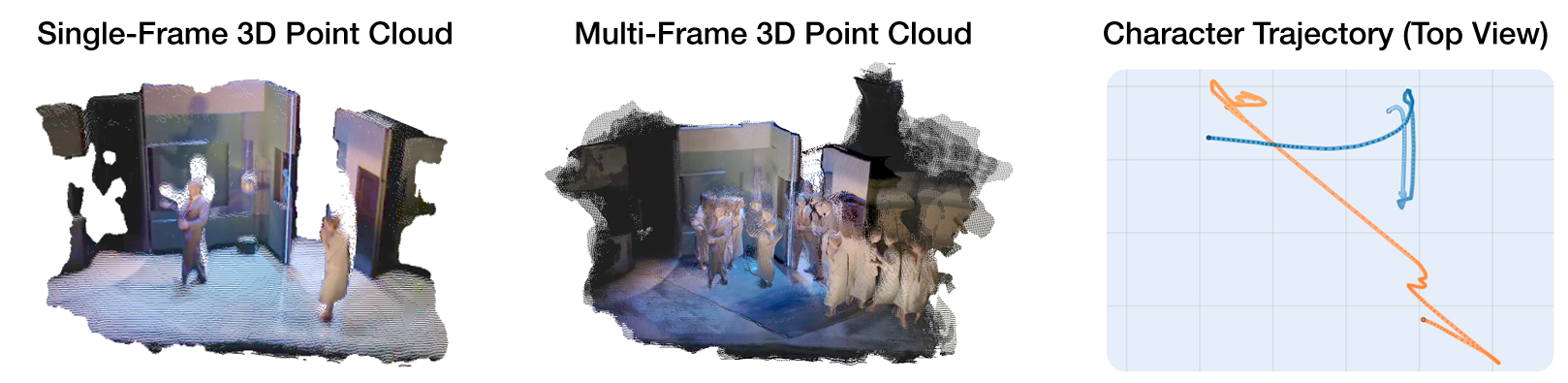}
    \caption{An example of the 3D point cloud and character trajectories reconstructed by \toole. The trajectory variations mostly arise from subtle body movements, such as swaying in place.}
    \label{fig:3D_reconstruction_details}
\end{figure*}

\end{document}